\documentclass[sigconf,nonacm]{acmart}

\usepackage{graphicx}

\setcopyright{rightsretained}
\copyrightyear{2023}

\begin{document}

    \title{A Moveable Beast:\\Partitioning Data and Compute for Computational Storage}

\author{Aldrin Montana}
\email{akmontan@ucsc.edu}
\orcid{0000-0003-2073-4813}
\affiliation{%
  \institution{UC Santa Cruz}
  \city{Santa Cruz}
  \state{CA}
  \country{USA}
}

\author{Yuanqing Xue}
\affiliation{%
  \institution{UC Santa Cruz}
  \city{Santa Cruz}
  \state{CA}
  \country{USA}
}

\author{Jeff LeFevre}
\affiliation{%
  \institution{UC Santa Cruz}
  \city{Santa Cruz}
  \state{CA}
  \country{USA}
}

\author{Carlos Maltzahn}
\affiliation{%
  \institution{UC Santa Cruz}
  \city{Santa Cruz}
  \state{CA}
  \country{USA}
}

\author{Josh Stuart}
\orcid{0000-0002-2171-565X}
\affiliation{%
  \institution{UC Santa Cruz}
  \city{Santa Cruz}
  \state{CA}
  \country{USA}
}

\author{Philip Kufeldt}
\affiliation{%
  \institution{Seagate Technology}
  \streetaddress{47488 Kato Rd}
  \city{Fremont}
  \state{CA}
  \country{USA}
}

\author{Peter Alvaro}
\affiliation{%
  \institution{UC Santa Cruz}
  \streetaddress{1156 High St}
  \city{Santa Cruz}
  \state{CA}
  \country{USA}
}

\authorsaddresses{}

\renewcommand{\shortauthors}{Montana, et al.}


    \begin{abstract}
        Over the years, hardware trends have introduced a variety of heterogeneous
        compute units while also bringing network and storage bandwidths within an order
        of magnitude of memory subsystems. In response, developers have used increasingly
        exotic solutions to extract more performance from hardware; typically relying on
        static, design-time partitioning of their programs which cannot keep pace with
        storage systems that are layering compute units throughout deepening hierarchies
        of storage devices.

        We argue that dynamic, just-in-time partitioning of computation offers a solution
        for emerging data systems to overcome ever-growing data sizes in the face of
        stalled CPU performance and memory bandwidth. In this paper, we describe our
        prototype computational storage system (CSS), \emph{Skytether}, that adopts a
        database perspective to utilize computational storage dri\-ves (CSDs). We also
        present \emph{MSG Express}, a data management system for single-cell gene
        expression data that sits on top of \textit{Skytether}. We discuss four design
        principles that guide the design of our CSS: support scientific applications;
        maximize utilization of storage, network, and memory bandwidth; minimize data
        movement; and enable flexible program execution on autonomous CSDs.
        \textit{Skytether} is designed for the extra layer of indirection that CSDs
        introduce to a storage system, using decomposable queries to take a new approach
        to computational storage that has been imagined but not yet explored.

        We use microbenchmarks to evaluate aspects of our initial progress: the impact of
        partition strategies, the relative cost of function execution on Kinetic drives,
        and the relative performance between two relational operators (selection and
        projection). Our use case measures differential expression, a comparison of
        quantified gene expression levels between two or more groups of biological cells.
        Based on processor clo\-ck rat\-es, we expected ~$3-4x$ performance slowdown on
        the computational storage engine (CSE) of Kinetic drives compared to a
        consumer-grade client CPU; instead, we observed an unexpected slowdown of ~$15x$.
        Fortunately, our evaluation results help us set anchor points in the design space
        for developing a cost model for decomposable queries and partitioning data across
        many CSDs.
    \end{abstract}

    \keywords{computational storage, storage, data management, gene express\-ion, single-cell}

    \maketitle



\section{Introduction} \label{sec:intro}
    For more than two decades, domain specialists who program data-intensive systems have had
    to reach for increasingly exotic solutions to extract more performance from hardware as
    data sizes inexorably grow. In the bygone days of Moore's law and Dennard scaling, these
    specialists could \emph{wait} for better performance; then, as CPU improvements slowed, they
    had to become experts in multicore parallelism as well as their primary domain. Partitioning a
    big data problem into roughly uniform pieces that can be processed in parallel while minimizing
    coordination remains a difficult open problem, far afield from domains such as data science,
    genomics, astronomy, and high-energy physics. Recent advances continue to exacerbate the issue
    with the introduction of a variety of heterogeneous compute units: computational storage
    drives, FPGAs, GPGPUs, TPUs, smart NICs, and DPUs. The confluence of hardware trends and
    growing data sizes requires that programmers not only partition their data as before, but must
    find a way to partition their \emph{program} to increase application or system performance.

    State-of-the-art solutions that take advantage of heterogeneous compute typically follow a
    static, design-time partitioning of a program. For computational storage drives (CSDs), a
    common appr\-oach has been to identify a computational ``kernel'' (e.g., a simple filter or
    transformation) that fits the device
    constraints~\cite{conf:smartssd-query,conf:yoursql,conf:polardb-csd}. Execution of the kernel
    can be ``pushed down'' to the CSD, a storage element containing one or more computational
    storage engines (CSE) and persistent data storage. This offloads work from the host's CPU (or
    CSE) and memory subsystem. While effective, this approach is fragile.

    An optimal, static partitioning of a program is likely to change whenever workload
    characteristics shift and for varying device characteristics. Additionally, bandwidths of
    network and storage devices have advanced to within one order of magnitude of memory bandwidth;
    shifting performance bottlenecks to the memory subsystem with a small number of these devices.
    As the gap between memory, network, and storage bandwidths shrink, compute units will continue
    to be layered throughout the storage hierarchy to keep pace with growing data sizes. Static
    partitioning of programs cannot keep up with layers of heterogeneous compute; a more general
    approach is necessary to partition and distribute programs effectively across deeper storage
    hierarchies.

    We argue that emerging data-intensive systems can be designed to overcome these limitations and
    support dynamic, just-in-time partitioning of computation across heterogeneous resources by
    applying a few well-known database concepts: the relational model, the notion of data
    independence, query planning and process\-ing, and optimization techniques. Our solution,
    ``decomposable quer\-ies,'' involves decomposing a query plan into a \textit{super-plan} and
    \textit{sub-plans} where each sub-plan is a complete, independent query plan. This approach
    requires a coordinated representation, of data and of the expressions which transform it, at
    each level of the storage hierarchy containing CSEs. A coordinated understanding of this
    representation enables the movement of data up the storage hierarchy, the movement of
    expressions down the storage hierarchy, or both.

    We draw our motivating use case from a biomolecular engineering research application
    involving the analysis of single-cell gene expression data. Piggybacking on
    advancements in DNA sequencing, single-cell technologies have revolutionized
    molecular biology. Where genomics can convey what versions of genes are present in a
    cell, single-cell RNA sequencing reveals what genes are transcribed, and possibly
    used, in a cell.  Biologists process single-cell transcriptomics data using various
    pipelines to produce \emph{single-cell gene expression} data, representing the
    quantities of each gene found in each cell. While there are international data
    repositories for single-cell gene expression, such as the \textit{Human Cell Atlas}
    or the \textit{EBI Gene Expression Atlas}, there are no existing efficient systems to
    support biologists in probing these datasets. Our motivating use case is to support
    identification of clusters within, and across, gene expression datasets. This
    requires providing biologists with support to transparently leverage modern,
    heterogeneous devices with minimal added complexity.

    In this paper, we describe our prototype computational storage system,
    \emph{Skytether}, that offers a solution to the crisis of ever-growing data sizes in
    the face of stalled CPU performance and memory bandwidth by adopting a database
    perspective. \textit{Skytether} is designed to partition data and program execution
    across a hierarchy of CSEs while minimizing CPU overheads and maximizing utilization
    of storage, network, and memory bandwidth. We also present \emph{MSG Express}, a data
    management system for single-cell gene expression data that sits on top of Skytether.
    Together, \textit{Skytether} and \textit{MSG Express} leverage existing technologies
    to transparently support analysis of gene expression levels for scientists who are
    accustomed to running jobs on their laptops over tiny data sizes. We quantify a
    measure of \textit{differential expression} using a \textit{T-statistic}.
    Differential expression reflects the levels of gene expression in one group of cells
    contrasted against another group and is a fundamental calculation for many scRNA-seq
    analyses. We show how gene expression data can be partitioned and how the
    \textit{T-Statistic} can be calculated using CSDs. We then present some experiments
    to evaluate the current generation of Seagate's Kinetic drives, using specific
    computation and filter pushdowns, to justify our decisions and inform future work
    (such as cost-based query optimization).

    Section~\ref{sec:scrna} discusses the urgency and the potential impact of data
    management for our data domain. In section~\ref{sec:background}, we discuss background
    and related work on smart drives and programmable storage that has influenced our
    approach of decomposable queries. In section~\ref{sec:design}, we discuss our design
    principles and decisions, present an experiment to justify our partition strategy for
    our data domain, and discuss how our design insulates us from specific hardware
    decisions (providing the benefits seen in bygone days). In
    section~\ref{sec:evaluation}, we present experiments to: ($1$) measure the performance
    of a particular computation kernel over a variety of hardware configurations using a
    single computational storage device and ($2$) an experiment to measure the performance
    of simple relational operations on a host CPU compared to a device CPU. We determine
    that the current generation of Kinetic drives do not yet provide efficient device-side
    query evaluation. In section~\ref{sec:discussion}, we reflect on the experimental
    results from section~\ref{sec:evaluation} and how they will inform future work on
    developing a cost-model for decomposable queries.

\section{Motivation} \label{sec:scrna}

    Current molecular biology and genomics approaches, especially pertaining to
    single-cell technologies, have a desperate need for more efficient and performant
    analytics solutions. Genomics has seen an explosion of data over the last 20 years.
    DNA sequencers can now produce raw data outputs from 60 GB to 360 GB to 16
    TB~\cite{web:pacbio-sequel,web:illumina-sequencers,web:ont-sequencers} and this trend
    is still continuing. However, DNA and RNA sequencing can only give a broad
    understanding of the genome and cellular state for an experiment as a whole.
    Single-cell technologies enable biologists to probe the genomes (DNA) and
    transcriptomes (RNA) for hundreds of thousands of individual cells in a single
    experiment, achieving unprecedented levels of resolution about tissue organization,
    organism development, and disease processes~\cite{conf:scrna-overview}.
    
    Single-cell RNA sequencing (\textit{scRNAseq}) has continued to improve and evolve
    since it was developed in 2014, enriching already complex data with even more
    structure, thus increasing the size of datasets (\textit{expression matrices}). More
    labs continue to adopt scRNAseq for their own research; consequently, the amount of
    scRNAseq data has grown exponentially and biologists now face a daunting challenge to
    compare experimental results with previously published results. Further compounding
    this daunting challenge, bioinformatics consortiums, such as the \textit{Human Cell
    Atlas} (\textit{HCA}), are serving as hubs for datasets and workloads from many
    international research labs.
    
    Single-cell gene expression data lends itself well to partitioning. Each single-cell
    experiment produces a matrix of data that can be easily encapsulated in its own
    dataset. Columns and rows of these datasets are independent and can be partitioned
    into multiple objects on a single storage device or across storage devices in
    straightforward ways. Although a particular data model and partition strategy can be
    straightforward, the various approaches greatly affect query performance; and, they
    are compounded by the variety of physical designs and query execution strategies.
    Then, especially for data housed by bioinformatics consortiums, biologists must
    perform extensive data integration and normalization. The design space and data
    processing requirements make it nearly impossible to effectively do scientific
    research, such as validate or determine the novelty of finding a new cell type or
    state, without becoming an expert in data management and storage systems. Empowering
    the management of single-cell gene expression data would better enable medical and/or
    biological insights such as discovering and characterizing the types of biological
    cells.
    
    \begin{figure}
    \centerline{\includegraphics[width=0.45\textwidth]{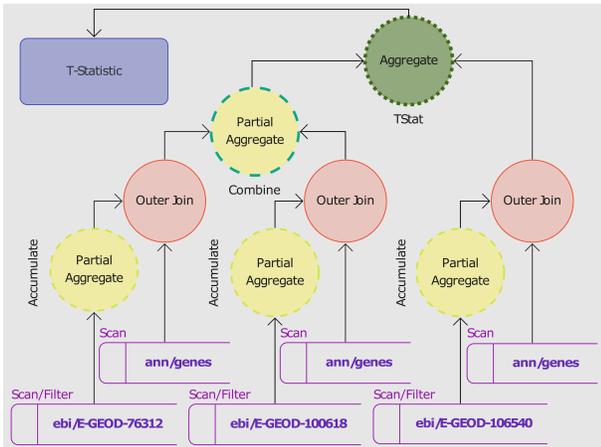}}
    \caption{Logical representation of differential expression (``T-statistic'') as a
             query plan. Actual implementation of the T-statistic calculation is
             imperative. A different colored border around a partial aggregate node
             denotes it to be a partial aggregate over multiple partial aggregates.
             Partial aggregates are computed over partition slices.}
    \label{fig:diff-expr-plan}
    \end{figure}

    \emph{Differential expression} quantifies the difference of gene expression levels
    between two or more groups of cells and underlies many analyses necessary to
    understand the role of various genes in normal and diseased settings. Thus, supporting
    efficient differential expression calculations presents a timely use case and is
    emblematic of a problem well-suited for computational storage. We use a
    \textit{T-statistic} to quantify differential expression, a statistical algorithm that
    can benefit from the same partitioning strategy as the gene expression data it is
    computed over. The \textit{T-statistic} is logically depicted as a query plan in
    Figure~\ref{fig:diff-expr-plan}.
    
    At its core, differential expression requires scanning the entirety of two input
    datasets, computing summary statistics, and then comparing those summaries. In some
    cases, and with various trade-offs, these datasets can be aggressively partitioned and
    their summaries can be computed incrementally.

\section{Background and Related Work} \label{sec:background}

    \textbf{Ceph and SkyhookDM.} We use Ceph and SkyhookDM as a starting point to design Skytether.
    Ceph is a distributed storage system that uses an object storage model and emphasizes
    \textit{reliability} and \textit{autonomy}~\cite{conf:ceph}. The goal of Ceph was to allow
    object storage devices (OSDs) in a storage cluster to act semi-autonomously while preserving
    consistency; maximizing availability; and performant, extensible data access~\cite{conf:rados}.
    Generally, an OSD is a storage service that uses the object storage model and runs on a server
    or ``intelligent device.'' Ceph also provides a powerful, \emph{extensible} data access
    interface that allows for application semantics to be defined closer to the data. SkyhookDM is
    a project that leverages this object interface to implement relational data access for Ceph
    storage objects~\cite{conf:skyhookdm,conf:skyhookdm-ceph,conf:skyhook-arrow}. Our prototype,
    Skytether, uses SkyhookDM as a primary reference for how to implement relational operations
    and query engine logic on OSDs; however, the design of Skytether is not tightly coupled to
    either implementation.

    In general, there are two key principles from Ceph that influence our design:
    \textit{autonomy} and \textit{extensibility}. Ceph achieves autonomy and scalability
    through \textit{shared nothing} data access and having an OSD store everything it
    needs to self-manage its state consistently. Ceph achieves extensibility through the
    use of object classes that can be registered with an OSD and customizes the logic
    executed on the access path for a data object. With autonomy as a guiding principle,
    we ensure that CSDs store everything they need to self-manage their state or execute
    pushdowns (a ``pushed down'' program). With extensibility as a guiding principle, we
    ensure that an OSD (or OSD-like service) is able to execute a program stored in the
    CSD. Shared nothing data access ensures that communication within the storage
    hierarchy is bounded eliminates dependencies between objects and allows for scale-out
    over many storage devices.

    For \textit{Skytether}, we assume access to Ceph or a Ceph-like system for scalability
    and reliability features such as replication. Then, to decouple OSD state and CSD
    state, we introduce a nested, independent key-value namespace for data stored on the
    CSD. This accommodates CSDs in a way that preserves the benefits of Ceph's
    architecture, ensuring that the OSD and CSD do not need to coordinate on data names.
    Because OSDs are strictly earlier on the data access path than CSDs, key-value names
    can be derived from object names (and are thus balanced and
    bounded~\cite{conf:tintenfisch}) and thus can be generated by any processor in the
    storage system. This makes coordination easy on the CSD side and hard on the OSD side.
    However, we note that CSDs are capable of storing aliases for data names locally for
    device-specific reasons further motivating this approach.

    In addition to the design principles we adopt, Ceph supports custom storage backends for
    efficient utilization of various storage devices. In May 2020, Aghayev, Weil, and others
    described research on ``BlueStore,'' a new storage backend for Ceph~\cite{journal:bluestore}.
    As part of their paper, the authors argue that new, custom storage backends can provide great
    benefits compared to fitting general-purpose file system abstractions to their needs. This
    world view naturally aligns with enabling autonomy of CSDs.

    
    \textbf{CSDs and Kinetic Drives.} Computational storage drives (\emph{CSD}) have
    been around for decades. They were first explored as ``database machines'' in the
    1970s and 1980s~\cite{journal:relational-hw,conf:dbmachine}, then as ``intelligent
    disks'' in the late 1990s~\cite{conf:idisks,phd:arch-dbapps}. In the late 1990s and
    early 2000s, they were also called ``active
    disks''~\cite{conf:active-model,conf:active-eval,phd:active-nas}. As storage drives
    equipped with modest processors and working memory, they pres\-ent an alluring
    opportunity to improve data processing and retrieval performance by moving
    computational kernels to the data; an opportunity made more attractive now that CPUs
    can no longer prom\-ise exponential increases in performance over time and on-device
    bus bandwidths are better able to move data into on-device CPUs. We use CSDs as an
    approach to dispersing, or scaling out, available compute resources.

\begin{figure}
    \centerline{\includegraphics[width=0.4\textwidth]{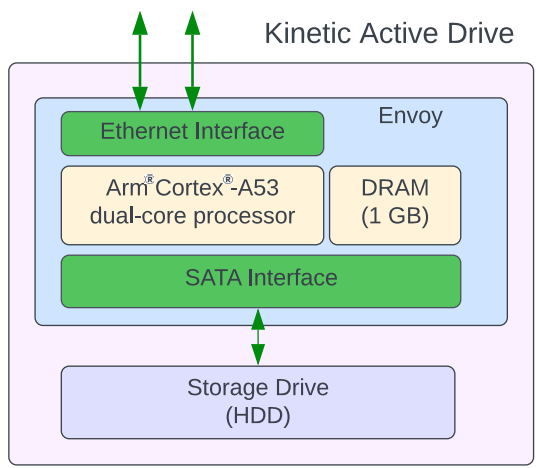}}
    \caption{Simplified hardware architecture of a Kinetic drive (CSD). The computational
             storage engine (CSE) is located on Envoy (the SoC module). Envoy
             communicates with a host through the ethernet interface and with persistent
             storage through the SATA interface. The current Kinetic drive implementation
             uses hard drives (HDD) for persistent storage.}
    \label{fig:kinetic}
    \end{figure}

    Due to the storage requirements of scRNAseq data, we specifically use CSDs with hard
    drives (HDDs) as persistent storage, as opposed to solid-state drives (SSDs) where
    much research has been done. Our prototype computational storage system (\emph{CSS})
    uses Kinetic drives--a Seagate research vehicle for an inexpensive, modular approach
    to computational storage. The current implementation of Kinetic drives uses a module,
    called \emph{Envoy}, containing a system-on-chip (SoC) as the computational storage
    engine (CSE) and a discrete HDD for persistent storage (depicted in
    Figure~\ref{fig:kinetic}). Data is accessed via the \emph{kinetic protocol} which
    provides a key-value interface~\cite{web:seagate-kinetic,code:kinetic-protocol}. The
    current Kinetic drive implementation trades modularity for performance: the additional
    cost and complexity of a Kinetic drive is concentrated on Envoy (and, subsequently,
    some accommodations by an enclosure).

    Envoy contains a general-purpose, power-efficient CPU and provides a familiar, server-like
    operating system that allows developers to: load shared libraries onto the device; store
    binaries in key-values as ``plain-old data''; and load programs from key-values, dynamically
    linking and executing them. Program execution looks as if we are executing it from the SoC
    directly. Programs can write results to the drive, but can also return results synchronously if
    desired.

    Architecturally, research on the ``Newport CSD'' from October 2020 most closely
    resembles Kinetic drives, with a similar architecture and execution
    environment~\cite{journal:newport}. The Newport CSD is described to have three
    distinct subsystems: (CS) the computing subsystem which, like a CSE, processes data;
    (FE) the front-end subsystem listens for NVMe commands and translates read and write
    commands to the back-end subsystem; (BE) the back-end subsystem that runs the storage
    controller logic such as wear levelling, garbage collection, and flash translation
    layer (FTL).

    The Newport CSD and current generation of Kinetic drives both use the same
    general-purpose Arm\textsuperscript{\copyright} processor
    (Arm\textsuperscript{\copyright}
    Cortex\textsuperscript{\copyright}-A53~\cite{web:arm-a53}) for the CS and a
    server-like OS. The Newport CSD has an ``Arm\textsuperscript{\copyright} M7'' for its
    FE and another one for its BE. Also, the BE shares silicon with the FE and CS. In
    comparison, the Envoy on a Kinetic drive is both the FE and CS; thus requiring that
    delegating compute to the Kinetic drive be carefully balanced with in-flight data
    accesses. Additionally, the BE for the Kinetic drive is a discrete storage device
    connected by a SATA interface, meaning that there is no shared silicon between the
    persistent storage and Envoy. However, the SATA interface is not a limiting factor for
    Kinetic drives which use an HDD for persistent storage.
    
    Although we use Kinetic drives in this paper, their similarities with Newport CSDs provide a
    perfect example of when decomposable queries can be effective. A disk service,
    \textit{Kinetic AD}, runs on Envoy and implements the server side of the kinetic protocol.
    \textit{Kinetic AD} decouples high-level data access (gets and puts) from low-level data
    persistence (writing and reading data from the block device). We use this key-value interface
    to access semantically meaningful dataset slices and to provide other high-level mechanisms
    while \textit{Kinetic AD} manages the storage device itself to optimally store and access
    key-values. To use a Newport CSD as we do Kinetic drives, we simply need to add a service,
    similar to \textit{Kinetic AD}, that insulates data access from the specific hardware
    architecture of the computational storage device. The only decision would be whether to run an
    object-level storage service or a key-value storage service on the Newport CSD.


    \textbf{Computational Storage System.} The approaches to computational storage most similar to
    ours are from the bodies of work about intelligent disks~\cite{phd:arch-dbapps} and active
    disks~\cite{phd:active-nas}. Research in both of these areas discuss CSDs with general-purpose
    CSEs and server-like operating systems running on each CSE. The similarities are unsurprising
    when considering that Kinetic drives are Seagate's research vehicle for a modern, spiritual
    successor to active disks. However, in both cases the researchers use the lens of a traditional,
    relational DBMS which differs from our approach leveraging an object storage model.

    Keeton mentions a software architecture similar to what we propose with Skytether on
    Kinetic drives: $1$--run a complete shared-nothing database server and operating
    system on each CSD~\cite{phd:arch-dbapps}.  However, her dissertation evaluates a
    hypothetical system and se\-em to use a different software architecture: $4$--run a
    reduced operating system, the storage/data manager, and relational operators on each
    CSD.

    Riedel mentions two design issues at a high level, but does not imple\-ment them or
    detail approaches: ($1$) parti\-tion\-ing of code for active disks and ($2$) why dynamic
    code~\cite{phd:active-nas}. What Rie\-del sugge\-sts for parti\-tion\-ing of code aligns
    very closely with decomposable quer\-ies and his definition of dynamic code aligns very
    closely with our idea of storing a query engine on a CSD to be loaded and exe\-cu\-ted on
    the CSE.


    Since 2012 there has been much research on computational storage using active
    flash~\cite{conf:active-outofcore,conf:active-insitu,conf:active-issds} and
    SSDs~\cite{conf:smartssd-cost,conf:smartssd-query,conf:ibex,conf:yoursql,
    journal:insitu-scansjoins,conf:blockndp,journal:newport,arxiv:zcsd,conf:csd-htap}. Despite the
    wealth of research in computational storage, we do not know of any \emph{recent} published work
    that aims to send query plans to a CSD for independent processing and execution. The closest we
    have found is Ibex~\cite{conf:ibex}--which supported \textit{selection}, \textit{projection},
    and \emph{GROUP BY} \textit{aggregation} as pipelined components in an FPGA--and research from
    Sungchan Kim, Hyunok Oh, et al.~\cite{journal:insitu-scansjoins}. Sungchan Kim, Hyunok Oh, et
    al. discuss in-storage processing (ISP) at a very low level, detailing hardware components and
    customizing hardware logic executed on a flash memory
    controller~\cite{journal:insitu-scansjoins}. Other research describe work at a similarly low
    level interacting with flash controllers, FPGAs, or hardware-specific implementations:
    YourSQL~\cite{conf:yoursql} and work from Wei Cao, et al. extending POLARDB~\cite{conf:polardb-csd}.
    Tobias Vincon, et al. also describe similar work on near-data processing (NDP) for HTAP
    workloads~\cite{conf:csd-htap}; but, their work and work on IS-HBase (in-storage computing for
    HBase)~\cite{journal:csd-hbase} look at accelerating key-value databases. Most other research
    on query processing over computational storage explore push-down of a single operator or of
    supporting functions~\cite{conf:smartssd-query,conf:csd-lists}. There is also some work on
    computational storage for zoned namespaces (ZNS) SSDs that discusses an open source approach
    using eBPF (extended Berkeley packet filter)~\cite{arxiv:zcsd}.

    Ultimately, Keeton and Riedel both envision an approach similar to ours, but discuss
    and evaluate approaches similar to recent computational storage research that pushes down
    single relational operators or supporting functions. Also, nearly all of the research on
    computational storage uses the lens that data is distributed across CSDs and functions pushed
    to the CSEs process a shard of the data. Our approach to computational storage attempts to
    treat each CSD as its own sub-database, capable of managing itself but expected to cooperate
    within a storage hierarchy.





    \section{Design Principles} \label{sec:principles}
    In this section, we discuss our guiding principles and our design requirements. We
    have mentioned several requirements for our computational storage system (CSS) in
    passing, but here we formalize them:
    \begin{description}
        \item[R1] Support biologists and application developers as transparently as
                  possible.
        \item[R2] Maximize utilization of storage, network, and memory ba\-nd\-width.
        \item[R3] Minimize data movement through the storage hierarchy.
        \item[R4] Enable flexibility to add new, heterogeneous compute uni\-ts in a storage
                  hierarchy; especially, CSDs.
    \end{description}


    \textbf{Supporting biologists.} Our primary, guiding principle is to transparently
    support biologists in their management and analysis of single-cell gene expression
    data (\emph{gene expression}). Biologists collect scRNAseq datasets in discrete
    experiments, which are then stored for later analysis and re-analysis to quantify gene
    expression. Analysis of gene expression can then later be analyzed and re-analyzed to
    gain insight into the state and function of individual biological cells
    (single-cells). As gene expression data continues to grow over time,  becoming far
    too large to be hosted on a single host, it requires the use of hard drives (HDDs) as
    the primary storage medium. Our CSS should accommodate HDDs in the storage hierarchy.

    To best support biologists, our CSS should allow biologists to process gene expression
    data using familiar interfaces and without them being prescriptive about the physical
    design of the data or the architecture of the system. We expect that gene expression
    analysis will be written in the R or Python programming languages, so our CSS should
    support intuitive interfaces to these languages.

    Gene expression data is processed by a bio\-informatics pipe\-line, produ\-cing
    di\-scr\-ete datasets--expression matrices (\emph{expr matrices}). Ea\-ch \textit{expr
    matrix} can be treated as a distinct storage object that can be loaded directly into a
    CSS and presents a natural boundary for a partition strategy. \textit{Expr matrices}
    may have the same data properties, but can vary greatly in their metadata and
    scientific context; thus, the only extra information our CSS should require to load an
    \textit{expr matrix} is metadata describing the columns (single-cells) and rows
    (genes) represented in the \textit{expr matrix}.

    Directly loading \textit{expr matrices} with minimal transformation lowers the barrier
    to data ingest and efficient analysis which is valuable for scientific computing.
    Research in ``NoDB,'' from Alagiannis, Idreos, Ailamaki, and
    others~\cite{conf:nodb,conf:here-datafiles}, highlights some of the needs and
    benefits. The similarities between our approach and NoDB is mentioned in more detail
    after we define our physical database design (section~\ref{sec:physical}).

    \textbf{Maximizing bandwidth utilization.} With the gap between memory, network, and
    storage bandwidths shrinking, CSEs provide more than just an extra compute unit.
    Scaling out memory bandwidth at CSEs and CSDs is cheaper than scaling up memory
    bandwidth at expensive compute complexes (compute nodes in a cluster). However,
    pushing down compute is incredibly latency sensitive, as bottlenecks in the storage
    hierarchy are cumulative across CSEs on the data path. On the other hand, executing
    compute can be worth the overhead if it reduces unnecessary data movement into compute
    complexes. A more general approach than static partitioning of programs is necessary
    to keep pace with diversifying hardware and deepening storage hierarchies. An
    effective CSS should be able to opportunistically push down compute, but quickly adapt
    if actual workloads prove too intensive for CSEs lower in the storage hierarchy.


    \textbf{Minimizing data movement.} A promise of CSDs is that data movement can be
    reduced throughout the storage hierarchy. The amount of data accessed at a storage
    device is invariant (with respect to how it is stored), but the fewer buses a data
    object traverses, the less energy it uses and the fewer CPU cycles it consumes.
    Additionally, specifically for HDDs, we use the principle that data should not be
    re-visited too frequently. When data is pulled off of the HDD and the CSE is done with
    it, we can loosely assume that it is cached somewhere higher in the storage hierarchy.

    \textbf{Flexibility and Autonomy.} Our design prioritizes flexibility and autonomy.
    Flexibility refers to program execution that can be deferred, when a CSE is
    overloaded, to an earlier CSE in the data access path (higher in the storage
    hierarchy). Autonomy refers to physical data design and program execution on a CSD
    according to its device characteristics. Designing for both flexibility and autonomy
    enables the use of new, distinct CSDs or CSEs and allows for the mix of compute unit
    types to change.

    To enable flexibility, query plans should support annotations, or some plan-level
    metadata, of what sub-plans have been executed. For example, if a CSE determines that
    its load is too high, it can decide to pass data up the storage hierarchy and return
    an annotated query plan (marked as ``not executed''). The server, or upstream CSE, can
    see that the query plan was not executed by the downstream CSE and execute the plan on
    the incoming data. This provides flexibility for services with fewer resources or
    higher contention to execute fewer operations to reduce overall waiting. Additional
    information, such as quality-of-service metadata, can be included in the annotations;
    this could enable behaviors such as indicating when a push down can be attempted
    again.

    To enable autonomy (independence of CSEs), que\-ry pla\-ns shou\-ld be logi\-cal and
    expressed at a sufficiently high level. A CSE must be able to decompose a query plan,
    propagate \textit{sub-plans} down the storage hierarchy, and execute any remaining
    portion of the \textit{super-plan}. Additionally, the CSE should be able to optimize
    a received query plan with respect to its system characteristics and physical data
    design. Logical query plans communicate intent, but allow a CSE to decide how to
    satisfy that intent. In the case of CSEs with general-purpose processors, there may
    be minimal changes to the query plan. However, this approach provides the necessary
    indirection to allow CSEs with specialized accelerators to execute specialized
    functions or decompose the query plan for other CSEs or CSDs (directly attached or
    downstream in the storage hierarchy).

    Cost-based query optimization requires extensions of existing cost models to permit
    an optimizer to reason over different ``cuts'' of a query plan into upstream and
    downstream portions. From the perspective of a decomposable query system, a
    custom-built filter pushdown is a degenerate case of a cut, in which only the leaf of
    a query tree (an access method and a selection predicate) is evaluated on a
    downstream CSE; it would consider this plan among many others.

\section{Computational Storage System} \label{sec:design}

    The addition of computational storage devices (CSDs) to a storage system introduces an
    extra layer of indirection for compute--data accesses from a storage server can be in
    the form of programs and not just function or API calls. We call a storage system
    designed for this additional complexity a computational storage system (CSS). This
    section discusses our design for a CSS, which we view as many storage services running
    on computational storage engines (CSEs) in a storage hierarchy. At the bottom of the
    hierarchy are persistent storage components serving data, and throughout the hierarchy
    are CSEs that the data may pass through until it gets to a compute node (cluster or
    application). In many ways, a CSS can be thought of as a distributed DBMS over
    dis-aggregated, computational storage. We use this lens to take a new approach to
    computational storage that has been imagined but not yet explored.

    \begin{figure*}
    \centerline{\includegraphics[width=0.8\textwidth]{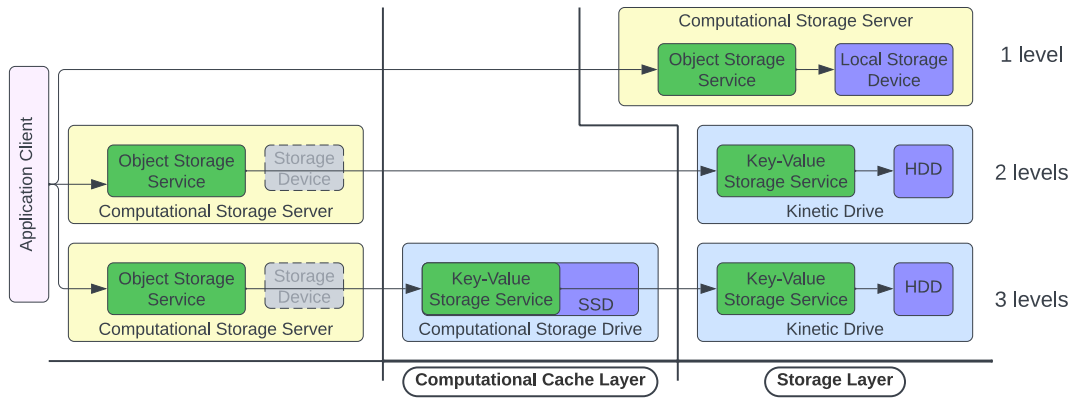}}
    \caption{High-level view of a computational storage system. Storage servers may be
             backed by a variety of storage devices (local or computational). A storage
             service handles requests and translates it into the appropriate back-end data
             access operation. If a Storage server is backed by a local storage device,
             only an object-level namespace is used. Kinetic drives house a discrete
             storage device, in contrast to a Newport CSD which is a single device.}
    \label{fig:cpuss-architecture}
    \end{figure*}


    \subsection{Physical Design} \label{sec:physical}

        For \textit{MSG Express}, a \textit{dataset} is a table containing application
        data that is handed to our storage system. Thus, datasets naturally represent an
        \textit{expr matrix}. We physically represent an \textit{expr matrix} as a table
        with the same layout (genes as rows), though the pivoted orientation can also be
        supported (genes as columns). For convenience, we support custom metadata attached
        to a \textit{dataset}; operations on \textit{datasets} propagate this metadata
        into the result \textit{dataset}. We also support application-defined groupings of
        \textit{datasets}, called \emph{domains}, to improve usability and as hints for
        performance. For example, we may want to have a domain for \textit{expr matrices}
        generated from a particular research lab or for a particular set of experiments
        that we expect to analyze together.

        \textbf{Storage Model.} Our data storage model has two primary layers: logical and
        physical. The logical layer, consisting of storage servers, uses an object storage
        model. The physical layer, consisting of CSDs, uses a key-value storage model. In
        general, these two models are similar in many ways, but in this paper we use
        objects to refer to data that may be decomposed and distributed, and key-values to
        refer to data that cannot be further decomposed. For example, in the case that
        there is only object-level storage, the underlying storage backend may still need
        to physically split the object's data.

        In contrast to other object storage systems, we map object locations to many
        storage servers to support parallel data access. We call this approach an
        \textit{autonomous object} model, where portions of an object may be managed by
        many storage services and each object can have independent (autonomous) physical
        design. A \textit{partition} refers to the portion of an object that is managed by
        a particular storage service. Autonomous physical design of \textit{partitions}
        can be leveraged for data that can be efficiently processed within a CSD, in
        which case transformation to a normalized physical design can be done on the CSD
        when returning results to a storage server, or it may be done on the storage
        server.

        \textbf{Data Model.} We split a \textit{dataset} into a set of \emph{partitions}
        that we \emph{may} distribute across many storage devices. Each \textit{partition}
        is split into many \emph{slices}, our smallest logical unit of storage. Some
        aspects of the physical design are established when splitting a
        \textit{partition}. These aspects are properties of the data and not stored as
        metadata; splitting can be done at any layer (even by the application) and will
        be respected throughout the CSS (to avoid unnecessary operations). There are many
        \emph{data slices} that contain the data and a single \emph{metadata slice}
        containing the schema for the partition and other metadata. Examples of
        \textit{slice} metadata include system-specific metadata such as indexes and
        physical design hints, or application-specific metadata which we transparently
        preserve.
        
        To ensure a loose coupling between the logical and physical layers, a
        \textit{slice} may be physically split across many key-values. This allows slicing
        of a \textit{partition} to occur at a higher level of abstraction as well as
        allowing a \textit{CSD} to alter the mapping of \textit{slices} to key-values for
        device-specific reasons. By default, a \textit{partition} is split across
        \textit{slices} such that a row is kept intact and a column is split. Then, we use
        the maximum size of a key-value to determine how many rows are contained in each
        \textit{slice}. If the \textit{slice} is striped across many key-values, then the
        total key-value size is used when maximizing \textit{slice} size. This decision is
        discussed further in Section~\ref{sec:eval-slicedims}.

        \textbf{Data Access Model.} As in other object storage systems, we store object
        names in a single namespace. Key-values are in CSD-local namespaces, meaning that
        the logical layer does not care about key names and if a \textit{partition} is not
        stored on a CSD, then data access goes through only the object-level namespace.
        Our system spans both logical and physical layers by naming \textit{slices} (by
        convention) using a dense, numerical suffix on the \textit{partition} key name.
        Using a simple convention means we can name any \textit{slice} from any CSD,
        obviating the need for the logical layer to manage \textit{slice} names. This
        naming also allows us to easily remember the order data was written (write-order)
        and index \textit{slices} by values of interest.

        When code is executed on the data access path for a particular object, other objects are
        inaccessible. In this way, objects have no logical dependence and so can be
        independently placed and replicated. Although, they can physically share resources
        by being managed by the same storage service. This design point is not enforceable
        at the CSD level unless we place constraints on the relationship between object
        names and key names. Although this seems like a drawback, it allows us to use
        \textit{metadata slices} for system-specific indirection to support features such
        as materialized views. Additionally, due to this relaxed constraint at the
        physical layer, we allow for object names to be remapped to new key-value names by
        the CSD.

        In general, the logical layer needs to balance partition load and partition
        utilization, whereas the physical layer needs to balance device load and device
        utilization. Partition load is how frequently a \textit{partition} is accessed.
        Partition utilization refers to the volume of data within a \textit{partition}
        that is frequently accessed, e.g. the relative (32\%) or absolute volume (24 GiB)
        or regional patterns (the first $x$ \textit{slices}).

        \textit{Slices} are an atomic unit for executing relational operations. If
        concurrent update requests conflict on a set of key-values, then one of the update
        requests will fail and none of its target key-values will be updated. This must be
        supported at the \textit{CSD} level, though it should suffice for it to be an
        in-memory mechanism if it is not supported in-storage. A \textit{slice} may also
        be called an \textit{in-memory slice} when it is in volatile memory or an
        \textit{in-storage slice} when we refer to how it is persisted on a storage
        device.

        The logical layer of our storage model handles \textit{partitions} and the
        physical layer handles \textit{slices}. In some ways, \textit{partitions} are
        akin to data pages and \textit{slices} are akin to blocks that a data page may be
        decomposed into. A \textit{domain} is comparable to a database schema (e.g.
        ``public''). These similarities allow us to accept and store datasets with
        minimal logical changes which provides benefits similar to NoDB~\cite{conf:nodb}.
        Practically, we are assuming that the extraction and transformation portions of
        ETL are handled by the application and the user-facing library that interacts
        with \textit{Skytether} (such as \textit{MSG Express}). This allows user
        applications to be more transparently accommodated by the CSS and the alignment
        of a database perspective with storage system concepts allows the CSS to better
        support database operations such as indexes and query processing.

        To maximize utilization of CSDs, our design prioritizes flexibility and autonomy.
        Flexibility refers to program execution that can be deferred, when a CSE is
        overloaded, to an earlier CSE in the data access path (higher in the storage
        hierarchy). Autonomy refers to physical data design and program execution on a CSD
        according to its device characteristics. Designing for both flexibility and autonomy
        enables the use of new, distinct CSDs or CSEs and allows for the mix of compute unit
        types to change.

    \subsection{Programming Model} \label{sec:dev}

        The \emph{Kinetic protocol}~\cite{code:kinetic-protocol} uses a key-value
        interface for data access and program execution. Program execution is initiated
        by an \emph{exec} command, which loads the program binary from a set of
        key-values and executes it. The executed program also uses the Kinetic protocol
        for data access; thus, a program that accesses data from a Kinetic drive can be
        stored and executed with almost no changes.

        To execute a query engine on a Kinetic drive, the binary for the query engine
        must be stored and executed. The \textit{libkinetic} library implements the
        kinetic protocol and can be used by the query engine for key-value data accesses.
        The \textit{exec} command accepts arguments and propagates them to the program
        being executed. For a query engine, arguments may include a query plan and other
        arguments to control behavior of the query engine.

        \textit{MSG Express} is primarily written in C++, but we support Python via
        Cython bindings to our C++ core. Then, we provide a convenience module that can
        be imported into R via the \textit{reticulate} package. In this way, we implement
        functionality in C++, provide useful Python bindings, and make it easy to use
        those Python bindings from R. Spanning these languages is how we coordinate both
        data representation and expressions throughout the application and storage
        hierarchy.

        To transparently support scientific applications and interface with an active data
        processing community, we use the Apache Arrow (\textit{Arrow}) library for data
        representation and expressing queries.
        Arrow supports bindings and popular data processing interfaces for both R and
        Python, enabling high-level applications with low overheads. Thus, we accept
        application data and expressions from either language, allowing our system to
        handle everything at the data management level.

    \subsection{Query Planning and Execution} \label{sec:queries}

\begin{figure}
        \centerline{\includegraphics[width=0.35\textwidth]{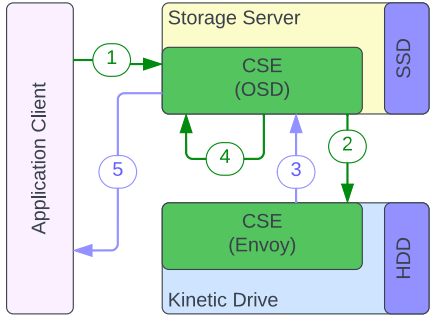}}
        \caption{High-level view of control and data flow for query processing in
                 \textit{Skytether}. $(1)$ send logical query plan to CSE, $(2)$ send
                 \textit{sub-plan} to downstream CSE, $(3)$ send results and annotated
                 query plan to upstream CSE, $(4)$ complete execution of the
                 \textit{super-plan} and remaining portions of the \textit{sub-plan},
                 $(5)$ send results to client. Results can be requested asynchronously if
                 preferred (steps $3$ and $5$ can be pull-based).}
        \label{fig:query-processing}
        \end{figure}

        \textit{MSG Express} takes expressions from R or Python, translates them to a
        query plan to be sent to a CSE running \textit{Skytether}. Then,
        \textit{Skyte\-ther} decomposes the query plan, propagates a \emph{sub-plan} to a
        downstream CSE, receives data and the same \textit{sub-plan} (with annotations),
        \emph{maybe} executes any of the query plan not executed, then propagates the
        result set upstream. Figure~\ref{fig:query-processing} shows a high-level
        overview of this sequence.

        We use substrait~\cite{web:substrait} to represent logical query plans and its
        annotations (in-tree or independent). Substrait provides an open, high-level query
        plan representation that we can optimize in two passes and execute in two passes.
        To maximize flexibility, optimization and execution may occur concurrently. The
        initial query plan is provided, or constructed, from user-facing libraries and so
        the query engine on each CSE only needs to parse and transform the substrait
        representation. Query optimization is naively done in two passes: ($1$) at an
        upstream CSE, such as a storage server at the logical layer, and ($2$) at a
        downstream CSE, such as a CSD at the physical layer. An optional third pass of
        optimization transforms the \textit{super-plan} for interactive, or incremental,
        execution concurrently on the upstream CSE (more details below).

        Optimization at an upstream CSE represents a best-effort ``request'' to a
        downstream CSE to process some data. This pass can be adaptive, requesting the
        storage device do more computation if it is able to execute the whole
        \textit{sub-plan} with minimal overhead; or, requesting the storage device do
        less computation if previously sent \textit{sub-plans} were only partially
        executed.

        Optimization at a downstream CSE represents a dynamic, real-time optimization of
        the query plan. The CSE will execute the query plan for some number of
        \textit{slices} to better predict the actual execution cost of the query plan. If the
        measured overhead is minimal, then the CSE may choose to process the entire query
        plan. If the measured overhead is above some threshold (which can be
        independently determined and adjusted), then the CSE may decide how much of the
        query plan to execute on the remaining \textit{slices}. In this case, the storage
        server may execute the remainder of the query plan itself.


        Query execution is done in two passes: ($1$) at a downstream CSE, and (2) at the
        upstream CSE. It is possible to receive the results synchronously, or retrieve
        results later asynchronously. This flexibility of a push or pull model of data
        movement allows us to overlap an optional third pass of optimization with both
        passes of execution. Also, due to each \textit{slice} requiring separate
        accesses, the query engine can choose to make result sets accessible after each
        \textit{slice}, some batch of \textit{slices}, or after all \textit{slices}.

        The simplest execution scenario occurs if the downstream CSE initiates \emph{push
        back} and does not execute any of the \textit{sub-plan}. Then, the upstream CSE
        will execute the remainder of the \textit{sub-plan} and the \textit{super-plan}
        on results from the downstream CSE. The most complex execution scenario is if the
        downstream CSE executes a portion of the \textit{sub-plan}, streams results and
        the annotated query plan back to the upstream CSE, and the upstream CSE executes
        the remainder of the \textit{sub-plan} and the \textit{super-plan} on the
        results--either as they arrive or in batches.

        The optional third pass of optimization is initiated when the storage device
        has completed execution of its \textit{sub-plan} on some initial \textit{slices}.
        In this case, the \textit{super-plan} may be merged with the incomplete portion
        of the \textit{sub-plan} and the merged query plan can be optimized. The decision
        for this additional, adaptive optimization pass can be made statically
        (configured) or dynamically (depending on query plan complexity such as how many
        relations to join).

        CSDs do not have infinite resources and are usually sized based on cost concerns.
        Therefore these devices may not have enough cycles for all requested compute. To
        effectively utilize network and storage bandwidth, it is necessary for program
        execution to be dynamic and adaptive, so that execution of a program portion can
        be ``pushed back'' up the storage hierarchy (deferred) to a more powerful CSE when
        load is high. In these cases, the accessed data would move up the hierarchy and
        the query plan annotated in a way that signifies that no work was done.


\section{Evaluation} \label{sec:evaluation}
    In this paper, we present an evaluation that focuses on the following three questions:
    \begin{description}
        \item[Exp 1] \textit{Differential expression} aggregates values within a row (a
                     gene), but these aggregates are best interpreted in groups (set of
                     genes). Should we co-locate more columns (single-cells) in a
                     \textit{slice}, or more rows (genes) in a \textit{slice}?

        \item[Exp 2] Pushing down compute is latency sensitive, as IO bottlenecks are
                     cumulative across CSEs on the data path. What is the code path
                     overhead of executing a query on Envoy via \textit{Kinetic AD}?

        \item[Exp 3] Two relational operators, \emph{selection} and \emph{projection},
                     are the simplest query plans we can push down to a CSD that have the
                     highest potential for reduced data movement. Ho\-w should we quantify
                     their cost?
    \end{description}


    For each experiment, we measure the performance (using latency in milliseconds) of
    some portion of calculating the differential expression t-statistic. As described in
    section~\ref{sec:scrna}, we use a t-statistic to measure differential expression
    between two datasets. For this paper, we implemented Student's formulation of
    t-statistic in C++ using Arrow. We use three functions to compute the different
    aggregations in Figure~\ref{fig:diff-expr-plan}. The first partial aggregate, applied
    after a selection and projection, is represented by \emph{Accumulate}. The second
    partial aggregate is represented by \emph{Combine} and merges two sets of partial
    aggregates (each the result of an \textit{Accumulate}) into a single set of partial
    aggregates. The final aggregate is represented by \emph{TStat} and merges two sets of
    partial aggregates into the final t-statistic result. These experiments cover a
    variety of performance characteristics and some initial experimental variables. Given
    the large design space, and the difficulty in initially setting anchor points within
    that design space, we leave further experiments for future work.

    \subsection{Experimental Hardware} \label{sec:hardware}
        For experimental hardware, we are interested in the processors and hard drive
        characteristics. Other components will be important for a full end-to-end
        evaluation in the future, but for our experiments in this paper we isolate
        processor and hard drive performance as much as possible.

        \textbf{Processors.} For the server, we used a consumer-grade \textit{x86-64} CPU:
        \textit{E3-1270 v3} (released in June 2013). This CPU has a base frequency of $3.5GHz$, $4$
        cores and $8$ threads, and cache sizes of $64KB$, $256KB$, and $8MB$ for $L1$, $L2$, and
        $L3$ caches, respectively.

        The CSD has a module, called \emph{Envoy}, which contains a
        Marvell\textsuperscript{\copyright} ARMADA 88F3720~\cite{web:espressobin-armada}
        system-on-chip (SoC). The SoC uses an Arm\textsuperscript{\copyright} v8-A CPU:
        \textit{Arm\textsuperscript{\copyright} Cortex\textsuperscript{\copyright}-A53}
        (released in October 2012). This CPU has a base frequency of $1GHz$ (up to
        $1.2GHz$), $2$ cores, and cache sizes of $32KB$ and $256KB$ for $L1$ and $L2$
        caches, respectively.


        \textbf{Hard Drives.} For the server, we used a consumer-grade HDD:
        \textit{ST2000DM008}. This is a 2TB SATA drive with $4096$ bytes per sector, $16$
        read/write heads, a cache buffer of $256MB$, and a maximum data transfer rate of
        $220 MB/s$.

        For the CSD, we used a nearline-grade HDD: \textit{ST16000NM000G}. This is a 16TB
        SATA drive with $4096$ bytes per sector, $18$ read/write heads, a cache buffer of
        $256MB$, and a maximum data transfer rate of $261 MB/s$.

    \subsection{Exp 1. Varying Slice Dimensions} \label{sec:eval-slicedims}

        \textbf{Motivation.} To take advantage of additional memory bandwidth in a
        computational storage hierarchy, data must be partitioned effectively. Analysis
        of gene expression data frequently filters on both rows and columns; but, we
        prefer to split an \textit{expr matrix} such that each \textit{slice} contains
        all of the columns (single-cells) for some subset of the rows (genes).

        For our data model (single-cells as columns, genes as rows), differential
        expression compares two sets of columns, representing two clusters (groups) of
        single-cells. Many approaches for measuring differential expression lend
        themselves towards a uniform partitioning of single-cell gene expression data;
        this is especially true for the \emph{T-Statistic}. Expression levels are
        group\-ed by row and aggregated into three summary statistics: mean, variance,
        cardinality. If a row is distributed amongst many CSDs, its data must travel up
        the storage hierarchy to calculate the summary statistics. If a column is
        distributed amongst many CSDs, then the summary statistics must travel up the
        storage hierarchy to determine the differential expression.

        Given the characteristics of differential expression, we expect that maximizing
        the width of slices (column count) will minimize data movement ($R3$). Access to
        many columns in a \textit{slice} maximizes temporal locality when aggregating
        values into summary statistics, benefiting from vector instructions to compute
        row-wise aggregations and cache utilization. Whereas, maximizing the height
        of slices (row count) will maximize bandwidth utilization ($R2$). Access to many
        rows in a \textit{slice} maximizes the number of CSDs a \textit{dataset} can be
        partitioned across and minimizes the number of accesses to load a whole column
        into memory from a \textit{partition}; thus, increasing parallelism of data
        access (across \textit{partitions}) and spatial locality (across \textit{slices}
        of a \textit{partition}).

        \begin{figure}
            \centerline{\includegraphics[width=0.5\textwidth]{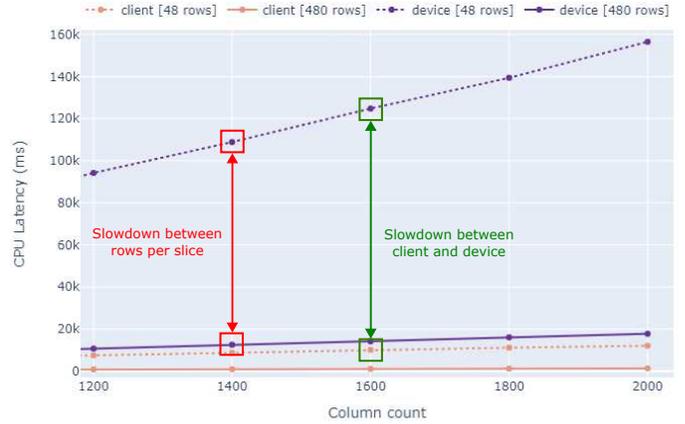}}
            \caption{The total CPU latency of executing \textit{Accumulate} on each slice
                     of a single \textit{dataset}. We measure latency directly on the CSE
                     and omit IO latency (\textit{Accumulate} is never waiting on IO).
                     Despite the expected impact of \textit{slice} dimensions, we find
                     that slice height has almost no effect on performance, only the
                     number of function invocations.}
            \label{fig:latency-by-dataset}
        \end{figure}

        \textbf{Setup.} We measure the total performance of executing \textit{Accumulate} on each
        slice of a single dataset. We apply the function on each slice of a partition, measuring
        latency of the function directly on the CSE and after the slice has been \emph{scanned} (no
        filtering) into memory so as to omit performance of the inter-device data path (HDD,
        networks) from the results. The slice characteristics we vary are: ($1$) the number of
        columns and ($2$) the number of rows. Slices are located on a single HDD, representing one
        of two scenarios: ($1$) a query plan has been received and can be fully satisfied within
        the CSD or ($2$) every sub-plan was pushed back and the CSE must execute this sub-plan over
        the returned slices.

        Figure~\ref{fig:latency-by-dataset} shows the total latency of executing
        \textit{Accumulate} on $300$ slices of a single partition
        (E-GEOD-76312~\cite{conf:scrna-76312,data:geod-76312}) to measure the sensitivity of a
        single partial aggregate to various slice dimensions (row and column counts). For clarity
        of reading the figure, there are four trend lines representing 4 combinations of CSE and
        row count. The trend lines, from top to bottom, are:
        \begin{description}
            \item[$1$; purple-dotted] $48$ rows on Envoy
            \item[$2$; purple-solid] $480$ rows on Envoy
            \item[$3$; orange-dotted] $48$ rows on client
            \item[$4$; orange-solid] $480$ rows on client
        \end{description}

        \textbf{Analysis.} To answer the question of sensitivity to slice dimensions, we compare
        trend lines $1$ and $2$, annotated in red on figure~\ref{fig:latency-by-dataset}. For a
        given slice height, increasing the slice width results in a minimal increase in latency.
        This validates that, for \textit{Accumulate}, very wide slices do not introduce a
        performance penalty. However, contrary to expectations, having fewer rows per slice seems
        to have a \emph{dramatic} performance penalty. We believe the performance penalty
        comes from the overhead of invoking \textit{Accumulate} many times and not from a
        difference in the different slice dimensions. The only difference between trend
        lines $1$ and $2$ is the height of each slice and invocations of API-level
        functions; the same amount of data is processed in both cases and the same number
        of arithmetic instructions are executed. This means that trend line $1$ invokes
        \textit{Accumulate} $300$ times, whereas trend line $2$ invokes
        \textit{Accumulate} $30$ times. We find that slice height also has almost no
        effect on performance, only the number of function invocations.

        In comparison to trend lines $1$ and $2$, trend lines $3$ and $4$ seem to have a much
        smaller gap in performance. That gap is better understood when we compare trend lines for
        the same slice dimensions but different CSEs. When we compare trend lines $1$ and $3$, we
        see that for slices of the same dimensions, the Envoy processor is approximately $13x$
        slower than the client processor (on average). The green annotations on the figure
        highlight this for slices with $48$ rows and $1600$ columns. Trend lines $2$ and $4$ show
        the same performance difference.

        \textbf{Takeaway.} The effects of row count and column count on total latency suggest that
        either partitioning strategy--vertical (columns) or horizontal (rows)--is viable from a
        processing perspective on a single CSD. However, the best partitioning strategy will be the
        one that results in fewer slices; or, many slices should be loaded into memory before
        executing a function such as \textit{Accumulate}.

    \subsection{Exp 2. Varying Execution Configurations}
        \textbf{Motivation.} To reduce bottlenecks at downstream CSDs and help improve
        bandwidth utilization, an extracted \textit{sub-plan} should be appropriately
        sized for the downstream CSD it will be sent to. The overhead of determining the
        portion of the \textit{sub-plan} to execute at a CSD is worthwhile if that cost
        is negligible or if the upstream and downstream CSDs can be confident in the
        sizing of the \textit{sub-plan}. Being able to mark a \textit{sub-plan} as being
        ``confidently sized'' would potentially create a fast-path for query optimization
        at the downstream CSD.

        To minimize bottlenecks ($R2$) when sending \textit{sub-plans} to a downstream
        CSD, we need to understand the code path overhead of executing a query plan on
        the CSD. Then, the overhead can be accommodated in a cost model to better
        size extracted \textit{sub-plans} to propagate downstream. To learn the code path
        overhead of \textit{Kinetic AD} and Kinetic drives, we compare a client processor
        to Envoy in a variety of configurations.

        Program execution on a Kinetic drive is initiated by sending a command,
        \textit{Exec}, to \textit{Kinetic AD} that specifies a key name prefix and
        arguments to propagate to the executed program. The key name prefix is used to
        load many keys into memory that collectively contain the program binary. Thus, a
        query is executed as follows:
        \begin{enumerate}
            \item Load binary for query engine
            \item Execute query engine with provided arguments (e.g. query string or query plan)
            \item Return status
        \end{enumerate}
        The query engine is like any C++ program that takes some arguments, executes some logic,
        then returns an output or writes the output to one or more key-values.

        \begin{figure*}
        \centerline{\includegraphics[width=\textwidth]{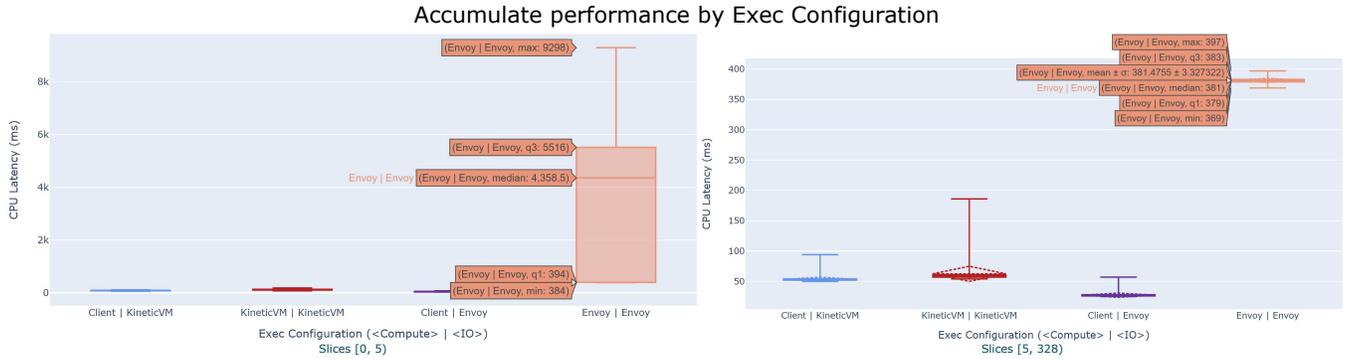}}
        \caption{CPU latency for each slice of a dataset in varying execution configurations.
                 An execution environment is labeled \textit{<Compute>$ | $<IO>}. The first portion
                 describes where a pushdown function is executed. The second portion describes what
                 storage device is used for data access.}
        \label{fig:latency-by-slice}
        \end{figure*}


        \textbf{Setup.} We use a specific $partition$ (E-MTAB-6819~\cite{data:mtab-6819}),
        measuring only the performance of executing \textit{Accumulate} on each slice
        (figure~\ref{fig:diff-expr-plan}) and plot summary statistics as a box plot.
        \textit{Accumulate} is implemented as a program that is stored on the Kinetic drive (using
        a \textit{put} command) instead of being passed as a query plan to a query engine.

        We use $4$ distinct \emph{execution configurations} (\textit{exec config}) to
        compare the relative performance of a client processor and the Envoy processor,
        and to control for the relative performance of a consumer-grade HDD (locally
        connected via SATA) and a nearline-grade HDD (connected to Envoy via SATA). An
        execution configuration is labeled as \textit{<Compute>$ | $<IO>}, where
        \textit{<Compute>} represents the processor executing the \textit{TStat} function:
        \begin{description}
            \item[\textit{Client}] \textit{x86-64} CPU on a host machine (ArchLinux)
            \item[\textit{KineticVM}] \textit{x86-64} CPU on a VirtualBox VM (Ubuntu)
            \item[\textit{Envoy}] Arm\textsuperscript{\copyright} v8-A CPU on the Envoy CSE
        \end{description}
        The second portion of the execution configuration, \textit{<IO>}, represents which HDD was
        used:
        \begin{description}
            \item[\textit{KineticVM}] consumer-grade HDD locally connected to the ho\-st
                                      machine but only accessed through the VM via
                                      ``VirtualBox raw vmdk.''
            \item[\textit{Envoy}] nearline-grade HDD directly connected to Envoy.
        \end{description}

        Each of the $4$ statistical summaries in figure~\ref{fig:latency-by-slice} show
        \textit{mean}, \textit{standard deviation}, \textit{min}, \textit{max}, \textit{lower
        quartile}, \textit{median}, and \textit{upper quartile}. Dashed lines show the mean and
        standard deviation as a diamond. Solid lines show the other statistics as a box
        plot. For conciseness, we number the configurations from left to right: the first
        configuration is \textit{Client$ | $KineticVM} and the fourth configuration is
        \textit{Envoy$ | $Envoy}.

        The \textit{Envoy$ | $Envoy}  configuration (Config$_1$) has significant outliers
        in the first $5$ slices of the partition, so we plot the first $5$ slices of each
        configuration separately without annotations for mean and standard deviation. In
        our experimental results, the compute thread was never waiting on a data request,
        so we omit timings for each slice data request from the figure.

        \textbf{Analysis.} We frame our analysis using $3$ pairs of execution configurations from
        Figure~\ref{fig:latency-by-slice}, which we number like so:
        \begin{enumerate}
            \item Config$_1$ and Config$_2$ (control)
            \item Config$_1$ and Config$_3$ (hard drive)
            \item Config$_4$ and Config$_3$ (compute location)
        \end{enumerate}
        The first configuration pair acts as a control, both the \textit{Client} and
        \textit{KineticVM} use the same hardware (CPU and HDD) and access data through
        \textit{Kinetic AD}. Additionally, the KineticVM executes the binary in the same
        way as on the \textit{Client} with only the extra overhead introduced by the
        virtual machine. We see that the virtual machine itself introduces some
        negligible slowdown. The second configuration pair highlights the effects of
        using a different HDD for data access. We see that the Envoy HDD (the HDD it
        uses) does not introduce any slowdown at all. The lower latencies are expected due
        to accessing the consumer-grade HDD via VirtualBox and the difference in
        performance between the consumer-grade HDD and nearline-grade HDD. Finally, the
        third configuration pair shows that the relative slowdown when running the compute
        function on the Envoy CPU is significant: ~$15x$. From the differences in base
        frequency and cache properties of the client and Envoy CPUs, we expect a
        difference of ~$3-4x$. This means that there is an additional ~$4-5x$ slowdown
        that comes from architectural differences. This means that, assuming perfect
        scale-out, $8$ Kinetic drives would have equivalent throughput as a single
        \textit{x86-64} CPU instead of the expected $4$ Kinetic drives.

        \textbf{Takeaway.} The code path overhead of executing a query on a Kinetic drive
        is much larger than expected. We expected an overhead of ~$3-4x$ but saw an
        overhead of ~$15x$. We believe much of this is specific to the pipeline
        architecture of the Envoy CPU (\textit{Arm\textsuperscript{\copyright}
        Cortex\textsuperscript{\copyright}-A53}) and the performance can be addressed. We
        also find that \textit{Kinetic AD} does not introduce much overhead.

        In section~\ref{sec:queries}, we mention that query optimization can use some
        initial \textit{slices} to determine actual costs of operations in the query
        plan. This experiment shows that for some CSEs (such as Envoy), initial slices
        may have abnormally high latency costs (left graph in
        Figure~\ref{fig:latency-by-slice}). For real-time optimization to be useful,
        \textit{Skytether} will need some mechanism to accommodate this discrepancy in
        performance between earlier and later \textit{slices}.

    \subsection{Exp 3. Latency of Selection and Projection}

        \textbf{Motivation.} To model cost for query plans we push down to CSDs, we want
        to understand the relative performance of relational operations. This
        understanding will allow us to assign costs to relational operations in a query
        plan, enabling decomposing and transforming query plans as appropriate for our
        data model and various CSE architectures. Accommodating costs of relational
        operations can be done in the query engine running on the CSD to support
        real-time optimization ($R4$) or it can be done in an upstream CSE if there is
        a sufficiently accurate understanding of the performance and capabilities of the
        downstream CSE ($R2$).

        \begin{figure}
            \centerline{\includegraphics[width=0.5\textwidth]{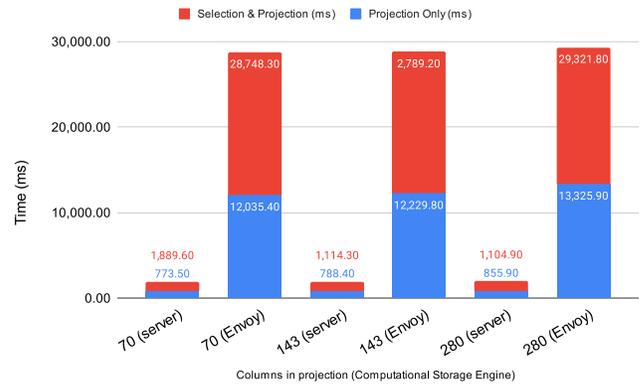}}
            \caption{CPU latency (in milliseconds) of executing selection and projection on a
                     partition (E-GEOD-76312). Selection predicate has ~$16.9\%$ selectivity,
                     projection sizes are for $3\%$, $7\%$, and $14\%$ of the partition width.}
            \label{fig:relops-latency}
        \end{figure}

        \textbf{Setup.} We evaluate relative performance of two relational operations that we
        expect to be the most common operations to push down to CSDs: \textit{selection} and
        \textit{projection}. We run this experiment on an Arrow table with $2,000$ columns and
        $12,000$ rows, which fits entirely into the Envoy's DRAM. We begin timing after the table
        is already loaded into memory to avoid HDD performance from obfuscating CPU performance.

        The \textit{selection} operation filters table rows using some predicate. For this
        experiment, we use a single inequality using an integer literal (\textit{SRR3052220}
        $> 10$) that represents a common use case to remove a particular aspect of noise in the
        data domain. This predicate has ~$16.9\%$ selectivity, meaning that ~$83\%$ of the rows are
        filtered out.

        The \textit{projection} operation filters table columns using some identifier. We vary the
        number of columns in the projection to be $3\%$, $7\%$, and $14\%$ of the partition's total
        columns. These column counts represent a projection of $1$, $2$, and $3$ column families,
        respectively, in the data domain (clusters of single-cells), and were chosen to verify that
        projection is not sensitive to column counts using semantically relevant values. Although
        projection is a simple operation, it is a higher-level function in the Arrow API than what
        is used in Experiment 1 to vary slice widths.

        Figure~\ref{fig:relops-latency} shows our experimental results, measuring average latency
        (in milliseconds; averaged over $10$ runs) of selection and projection as six stacked bars.
        Each bar represents the total average latency of projection and selection (red). The bottom
        portion of each bar (blue) is the average latency of only projection. The average latency
        for only the selection operation is not shown, but can be derived by subtracting the blue
        portion (lower number) from the red portion (higher number).

        \textbf{Analysis.} The stacked bars highlight that the projection operation is ~$40\%$ of
        the average latency of executing selection and projection. The relative timings of these
        operations are consistent for each projection size and each CSE, which suggests that
        neither impact the relative performance of projection.

        When comparing bars for different CSEs and for a particular projection size (e.g. $70$
        columns), we see that the Envoy processor has a significant slowdown of ~$15x$ in
        comparison to the server processor. This re-affirms the takeaway from Experiment 2 (but for
        higher-level functions) that the Envoy processor has an additional ~$4-5x$ slowdown than
        expected.

        \textbf{Takeaway.} The relative performance of selection and projection is that projection
        is faster by a small amount, compared to a selection predicate that uses a single column.
        This informs us that these two operations can be given a similar cost for the same size
        partition. We believe this costing generalizes to CSEs using either of these processor
        architectures (\textit{x86-64} and Arm\textsuperscript{\copyright} v8-A); despite the
        performance gap between the two CSEs, the relative performance between selection and
        projection is consistent and suggests to us that performance improvements for one operation
        will benefit the other operation equally.





    \section{Discussion and Future Work} \label{sec:discussion}
    We now have insight into the impact of partition strategies. It is surprising that
    both partition strategies seem to have similar performance on a single device for a
    straightforward use case. \emph{Experiment 1} purposely focused on omitting
    communication costs due to the variety of possible communication patterns.

    There are still significant trade-offs at higher levels of abstraction (the
    \textit{TStat} function vs the \textit{Accumulate} function), but this result means
    that various physical designs may be viable in combination with various communication
    scenarios. Future evaluation will accommodate this sizable design space to determine
    a cost model for a physical design and its impact on bandwidth utilization ($R2$) and
    data movement ($R3$).

    Using microbenchmarks, we have evaluated the relative cost of function execution on
    Envoy and observed an unexpected slowdown of ~$15x$ (instead of ~$3-4x$) on Envoy
    (the Kinetic CSE) for compute functions such as \textit{Accumulate}. This slowdown is
    also observable for simple relational operations--\emph{projection} and
    \emph{selection}. Despite the shortcomings of the current implementation of Envoy,
    hardware improvements can be independently researched and we believe such
    improvements can transparently improve performance of program execution.


    We have gained experience with Arrow for coordinating representation and expressions
    across a hierarchy of CSEs and scientific applications. Arrow appears to be a good
    choice, with continuing improvements that we can benefit from. Open source tooling
    that prioritizes interoperability allows for the development stack to gain new
    features while having lower less development burden on hardware engineers.

    In addition to compute functions and relational operations, we can benefit from Arrow
    Flight--a recent component of Arrow. Fli\-ght uses similar underlying technology as the
    Kinetic protocol but reduces data copies between an application and network library.
    This will generally reduce latencies for the query engine on a CSE when communicating
    with upstream or downstream CSEs. Flight would be especially useful for program
    execution on a Kinetic drive, which uses a network library to communicate with
    \textit{Kinetic AD} for each read and write operation on \textit{slices}. We also
    plan to integrate with, and evaluate, substrait for query plan representation to
    fully realize decomposable queries.

    In the future, we will use many CSDs together to validate our expectation that ~$16$
    CSDs will allow \textit{Skytether} to have an end-to-end latency comparable to a
    client processor for functions such as \textit{Accumulate}. We will use decomposable
    queries to show how we maximize bandwidth utilization ($R2$) by scaling across all of
    the CSDs, while also aggressively caching summary statistics (results from
    \textit{Accumulate} and \textit{Combine}) to minimize data movement ($R3$). There is
    a large design space to consider for: utilization of HDD capacity for cached results,
    the effect of CSD load on query optimization, and how many useful \textit{sub-plans}
    can be extracted from a complex query plan (like in figure~\ref{fig:diff-expr-plan}).
    It will be a challenge, but we aim to explore major points in this design space using
    decomposable queries and autonomous CSDs ($R4$).

    The evaluation results (section~\ref{sec:evaluation}) help us set anchor points in
    the design space for a cost model. We can now move forward with partitioning data
    across many CSDs and developing a generalizable cost model for aggregation functions
    and relational operators for CSDs. We believe we can continue our approach using
    Seagate's research CSDs--Kinetic drives--due to the \textit{Kinetic AD} interface. We
    also look forward to exploring more complex storage hierarchies, potentially with
    more levels and heterogeneity of CSEs ($R4$).

    The characteristics of single-cell gene expression data and differential analysis
    align well with partitioning of data and compute. Gene expression matrices can be
    naturally encapsulated in an independent dataset, where columns and rows are
    independent. We believe these characteristics also exist for high-energy physics
    (HEP) datasets where particles and observations of particle state can be stored
    independently. Where bioinformatics consortiums, such as the HCA, present a
    datacenter-like environment, HEP also has international, multi-lab collaborations
    such as the European Council for Nuclear Research (CERN). In future work, we plan to
    generalize our work by reusing relevant parts of \textit{MSG Express}, or possibly
    generalizing \textit{MSG Express} itself.

\begin{acks}
  The authors would like to thank \emph{Seagate Technology} for supporting this research.
  This material is also based upon work supported by the \emph{National Science
  Foundation} under Grants TI-2229773 and CNS-1764102, and Cooperative Agreement
  OAC-1836650, and the \emph{Center for Research in Open Source Software}
  (CROSS)~\cite{web:cross}.
\end{acks}

    \bibliographystyle{ACM-Reference-Format}
    \bibliography{references}

\end{document}